\documentclass[twocolumn,preprintnumbers,amsmath,amssymb]{revtex4}
\usepackage{graphicx}% Include figure files
\usepackage{bm}% bold math
\usepackage{latexsym}
\usepackage{epstopdf}
\newcommand{\bra}[1]{\langle #1|}
\newcommand{\ket}[1]{|#1\rangle}

\newcommand{\ketup}{| \!\!\uparrow \rangle}
\newcommand{\ketdown}{| \!\! \downarrow \rangle}

\begin{document}

%double-spaced
%\baselineskip24pt

%\preprint{ICPS 2006}
\title{Detection of single electron spin resonance in a double quantum dot}

\author{F. H. L. Koppens}
 \email{f.h.l.koppens@tudelft.nl}
\altaffiliation{Kavli Institute of NanoScience Delft, P.O. Box 5046, 2600 GA Delft, The Netherlands}

\author{C. Buizert}
%\altaffiliation{Kavli Institute of NanoScience Delft, P.O. Box 5046, 2600 GA Delft, The Netherlands}

\author{I. T. Vink}

%\altaffiliation{Kavli Institute of NanoScience Delft, P.O. Box 5046, 2600 GA Delft, The Netherlands}

\author{K.C. Nowack}

%\altaffiliation{Kavli Institute of NanoScience Delft, P.O. Box 5046, 2600 GA Delft, The Netherlands}

\author{T. Meunier}

%\altaffiliation{Kavli Institute of NanoScience Delft, P.O. Box 5046, 2600 GA Delft, The Netherlands}

\author{L. P. Kouwenhoven}

%\altaffiliation{Kavli Institute of NanoScience Delft, P.O. Box 5046, 2600 GA Delft, The Netherlands}

\author{L. M. K. Vandersypen}

%\altaffiliation{Kavli Institute of NanoScience Delft, P.O. Box 5046, 2600 GA Delft, The Netherlands}

%\vspace*

%\date{\today}

\begin{abstract}
Spin-dependent transport measurements through a double quantum dot are a valuable tool for detecting both the coherent evolution of the spin state of
a single electron as well as the hybridization of two-electron spin states. In this paper, we discuss a model that describes the transport cycle in
this regime, including the effects of an oscillating magnetic field (causing electron spin resonance) and the effective nuclear fields on the spin
states in the two dots. We numerically calculate the current flow due to the induced spin flips via electron spin resonance and we study the detector
efficiency for a range of parameters. The experimental data are compared with the model and we find a reasonable agreement.
\end{abstract}

\maketitle

\subsection{Introduction}

Recently, coherent spin rotations of a single electron were demonstrated in a double quantum dot device \cite{koppensnature}. In this system,
spin-flips of an electron in the dot were induced via an oscillating magnetic field (electron spin resonance or ESR) and detected through a
spin-dependent transition of the electron to another dot, which already contained one additional electron. This detection scheme is an extension of
the proposal for ESR detection in a single quantum dot by Engel and Loss \cite{engelesr}. Briefly, the device can be operated (in a spin blockade
regime \cite{onoscience}) such that the electron in the left dot can only move to the right dot if a spin flip in one of the two dots is induced via
ESR. From the right dot, the electron exits to the right reservoir and another electron enters the left dot from the left reservoir. A continuous
repetition of this transition will result in a net current flow.

Compared to the single dot detection scheme \cite{engelesr}, using the double-dot as the detector has two major advantages. First, the experiment can
be performed at a lower static magnetic field and consequently with lower, technically less demanding, excitation frequencies. Second, the spin
detection is rather insensitive to unwanted oscillating electric fields, because the relevant dot levels can be positioned far from the Fermi
energies of the leads. These electric fields are unavoidably generated together with the oscillating magnetic field as well.

The drawback of the double-dot detector is that spin detection is based on the projection in the two-electron singlet-triplet basis, while the aim is
to detect single spin rotations. However, this detection is still possible because the electrons in the two dots experience different effective
nuclear fields. This is due to the hyperfine interaction of the electron spins with the (roughly $10^6$) nuclear spins in the host semiconductor
material of each quantum dot \cite{johnsonnature,koppensscience,braun,gammonprl,merkulov,khaetskii,coishnonmarkovian,onoprl}. In order to provide
more insight in this double-dot ESR detection scheme for single spin rotations, it is necessary to analyze the coherent evolution of the two-electron
spin states together with the transitions in the transport cycle.

In this paper, we discuss a model that describes the transport cycle in the spin blockade regime while including the coherent coupling between the
two dots, and the influence of the static and oscillating magnetic field together with the effective nuclear fields on the electron spin states. The
aim is to understand how effectively single spin resonance will affect the measured quantity in the experiment, namely the current flow in the spin
blockade regime. The organization of this paper is as follows. First, we will explain the transport cycle and the mechanism that causes spin
blockade. Next, we will briefly discuss the static system Hamiltonian and the mixing of the two-electron spin states by the effective nuclear field.
Then we add an oscillating magnetic field to this Hamiltonian, that forms -together with the double dot tunnelling processes- the basis of the rate
equations that describe how the density matrix of the two-electron spin states evolves in time. The current flow, calculated from the steady state
solution of the density operator, is then analyzed for different coherent coupling values, magnitudes of the oscillating magnetic field, in
combination with different effective nuclear fields in the two dots. This provides further insight in the optimal conditions for spin-flip detection
with a double quantum dot.

\subsection{Spin blockade }

\begin{figure}[b]

          \includegraphics[scale=1.0]{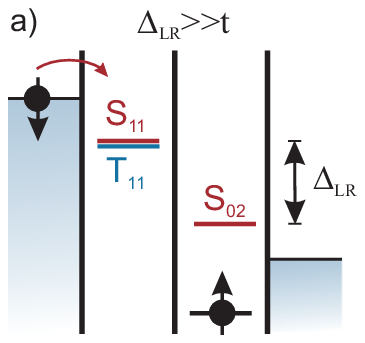}
     \includegraphics[scale=1.0]{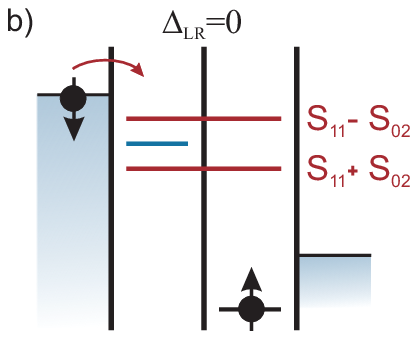}
          \includegraphics[scale=1.0]{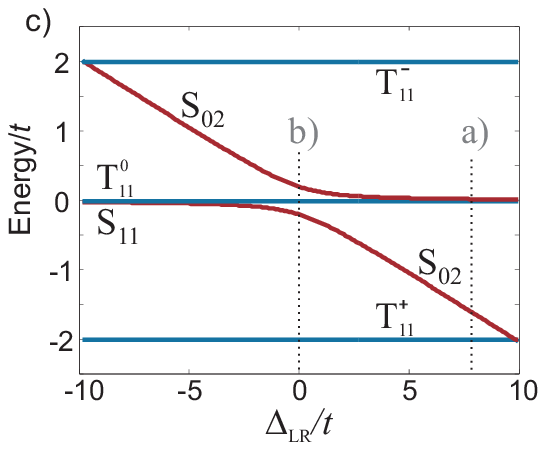}
   %  \centering
     \caption{a) A schematic of the double dot and the electro-chemical potentials (energy relative to the (0,1) state) of the relevant two-electron spin states. For $\Delta_{LR}>t$, transitions
     from the $S_{11}$ state to the $S_{02}$ state are possible via inelastic relaxation with rate
     $\Gamma_{in}$. Spin blockade occurs when one of the $T_{11}^i$ states is occupied. b) Similar schematic for $\Delta_{LR}=0$, where the singlet states are hybridized. Also in this case, spin blockade occurs when one of $T_{11}^i$ states is occupied.  c) Energy levels as a
     function of detuning. At $\Delta_{LR}=0$, the singlet states hybridize into bonding and anti-bonding states. The
     splitting between the triplets states corresponds to the Zeeman energy $g
     \mu_B B_{ext}$.}
     \label{F_DDlevels}
\end{figure}

In the spin-blockade regime, the double-dot is tuned such that one electron always resides in the right dot, and a second electron can tunnel from
the left reservoir through the left and right dots, to the right reservoir \cite{onoscience}. This current-carrying cycle can be described with the
occupations (m, n) of the left and right dots: $(1,1)\rightarrow (0,2)\rightarrow(0,1)\rightarrow(1,1)$. When an electron enters the left dot and
forms a double-dot singlet state $S_{11}$ with the electron in the right dot ($S=\ket{\!\uparrow\downarrow}- \ket{\!\downarrow\uparrow}$,
normalization omitted for brevity), it is possible for the left electron to move to the right dot, because the right dot singlet state $S_{02}$ is
energetically accessible. Next, one electron tunnels from the right dot to the right lead and another electron can again tunnel into the left dot.
If, however, the two electrons form a double-dot triplet state $T_{11}$, the left electron cannot move to the right dot, as the right dot triplet
state $T_{02}$ is much higher in energy (due to the relatively large exchange splitting in a single dot). The electron can also not move back to the
lead and therefore further current flow is blocked as soon as any of the (double-dot) triplet states is formed (Fig. \ref{F_DDlevels}a,b).

Spin blockade only occurs if at least one of the eigenstates of the system Hamiltonian is a pure triplet state. If processes are present that induce
transitions from all the three (1,1) triplet states to the (1,1) singlet state, spin blockade is lifted and a current will flow. As we will see
below, the presence of the nuclear spins in the host semiconductor can give rise to these kind of transitions. This can be seen most easily by adding
the effect of the hyperfine interaction to the system Hamiltonian.

\subsection{System Hamiltonian}

The system Hamiltonian is most conveniently written in the two-electron singlet-triplet basis with the quantization-axis in the z-direction. The
basis states are $S_{11}$, $T_{11}^+$, $T_{11}^-$, $T_{11}^0$ and $S_{02}$. The subscript $m,n$ denotes the dot occupancy. We exclude the $T_{02}$
state from the model, because this state is energetically inaccessible and therefore does not play an important role in the transport cycle.
Furthermore, we neglect the thermal energy $kT$ in the description, which is justified when the bias over the two dots is much larger than $kT$. The
system Hamiltonian is given by
\begin{eqnarray}
\label{E_WhatIzHamiltonian} H_{0} =&-&\Delta_{LR}\ket{S_{02}}\bra{S_{02}}+t\Bigl{(}\ket{S_{11}}\bra{S_{02}}+\ket{S_{02}}\bra{S_{11}} \Bigr{)}
\nonumber
\\
&-&g \mu_{B}B_{ext}\Bigl{(}\ket{T_{11}^+}\bra{T_{11}^+} - \ket{T_{11}^{-}}\bra{T_{11}^{-}}  \Bigr{)},
\end{eqnarray}
where $\Delta_{LR}$ is the energy difference between the $\ket{S_{11}}$ and $\ket{S_{02}}$ state (level detuning, see Fig.\ref{F_DDlevels}a), $t$ is
the tunnel coupling between the $\ket{S_{11}}$ and $\ket{S_{02}}$ states, $B_{ext}$ the external magnetic field in the z-direction and $S^{z}_{L(R)}$
the spin operator along $z$ for the left (right) electron. The eigenstates of the Hamiltonian (\ref{E_WhatIzHamiltonian}) for finite external field
are shown in figure \ref{F_DDlevels}c. For $\Delta_{LR}<t$, the tunnel coupling $t$ causes an anti-crossing of the $\ket{S_{11}}$ and $\ket{S_{02}}$
states. For $\Delta_{LR}<0$, transport is blocked by Coulomb blockade (i.e. the final state $\ket{S_{02}}$ is at a higher energy than the initial
state $S_{11}$). For $\Delta_{LR}\geq0$, transport will be blocked when one of the three triplet states becomes occupied (spin blockade). In
Fig.\ref{F_DDlevels}a and b, we distinguish two regimes: $\Delta_{LR}>t$ where the (exchange) energy splitting between $T_{11}^0$ and $S_{11}$ is
negligibly small and transitions from  $S_{11}$ to $S_{02}$ occur via inelastic relaxation with rate $\Gamma_{in}$ and the energy. A different regime
holds for $\Delta_{LR}<t$, where $S_{11}$ is coherently coupled with $S_{02}$ giving rise to a finite (exchange) splitting between $T_{11}^0$ and the
hybridized singlet states. We will return to this distinction in the discussion below.
%in order to put figures in text: change [p] to [h] and remove % before \includegraphics

\subsection{Singlet-triplet mixing by the nuclear spins}

\begin{figure}[h]
    \includegraphics[scale=1]{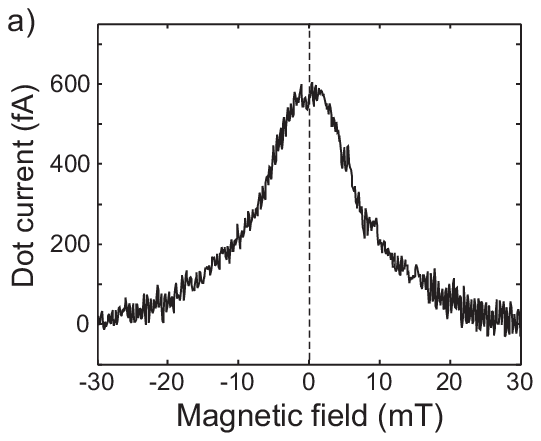}
          \includegraphics[scale=0.8]{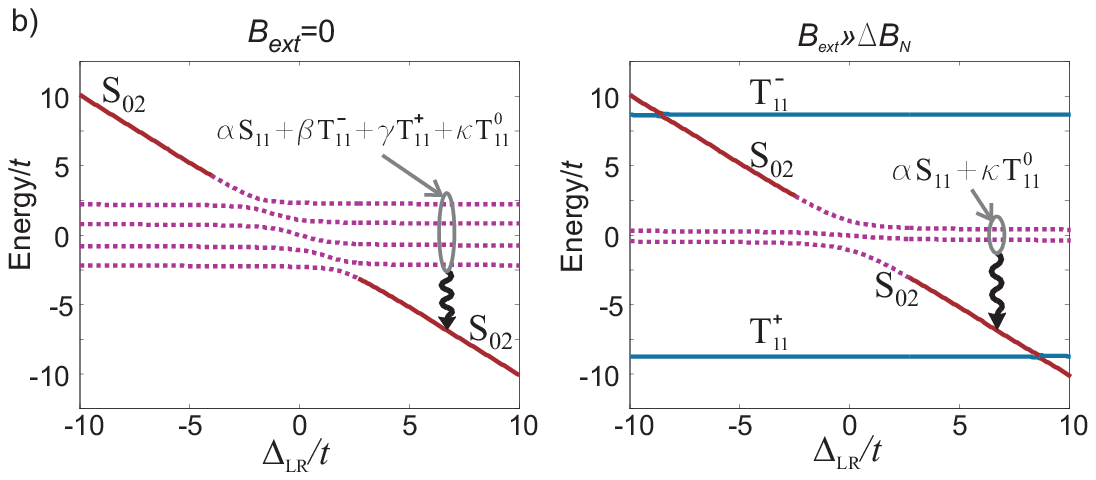}
     %\centering
     \caption{a) Observed current flow in the inelastic transport regime ($g\mu_B\Delta_{LR}\gg t$) due to singlet-triplet mixing by
     the nuclei. b) Electrochemical potentials in the presence of $H_{nucl}$ ($t\sim\Delta B_N$). Singlet and triplet eigenstates are denoted by red and blue lines respectively. Hybridized states (of singlet and triplet) are denoted by dotted purple lines. For $g\mu_BB_{ext}\gg t,g\mu_B\Delta {\bf B}_N$, the split-off
     triplets ($T^+_{11}$ and $T^-_{11}$) are hardly perturbed and current flow is blocked when they become occupied. Parameters: $t=0.2 \mu$eV, $g\mu_BB_{N,L}$=(0.1,0,-0.1)$\mu$eV, $g\mu_B B_{N,R}$=(-0.1,-0.2,-0.2)$\mu$eV and $g\mu_B B_{ext}$=2$\mu$eV.}
     \label{F_STmixing}
\end{figure}

The effect of the hyperfine interaction with the nuclear spins can be studied \cite{jouravlevprl} by adding a static (frozen) effective nuclear field
${\bf B}_{N}^{L}$ (${\bf B}_{N}^{R}$) at the left (right) dot to the system Hamiltonian:
\begin{eqnarray} \label{E_nuclpert}
H_{\rm nucl} &=& -g \mu_{B} \bigl{(} {\bf B}_{N}^{L} \cdot {\bf S}_{L} + {\bf B}_{N}^{R} \cdot {\bf S}_{R} \bigr{)} \nonumber
\\
&=& - g \mu_{B} ({\bf B}_{N}^{L} - {\bf B}_{N}^{R}) \cdot ({\bf S}_{L} - {\bf S}_{R})/2 \nonumber
\\
&&-g \mu_{B} ({\bf B}_{N}^{L}  + {\bf B}_{N}^{R}) \cdot ({\bf S}_{L} + {\bf S}_{R})/2.
\end{eqnarray}

For the sake of convenience, we separate the inhomogeneous and homogeneous contribution, for reasons which we will discuss later.  Considering the
nuclear field as static is justified since the tunnel rates and electron spin dynamics are expected to be much faster than the dynamics of the
nuclear system \cite{coishnonmarkovian,Sousa,Paget77}. Therefore, we will treat the Hamiltonian as time-independent. The effect of nuclear
reorientation will be included later by ensemble averaging.

We will show now that triplet states mix with the $S_{11}$ state if the nuclear field is different in the two dots (in all three directions). This
mixing will lift spin blockade, detectable as a finite current running through the dots for $\Delta_{LR}\geq0$. The effective nuclear field can be
decomposed in a homogeneous and an inhomogeneous part (see right-hand side of (\ref{E_nuclpert})). The homogeneous part simply adds vectorially to
the external field $B_{ext}$, changing slightly the Zeeman splitting and preferred spin orientation of the triplet states. The inhomogeneous part
$\Delta {\bf B}_N \equiv {\bf B}_N^L-{\bf B}_N^R $ on the other hand couples the triplet states to the singlet state, as can be seen readily by
combining the spin operators in the following way
\begin{eqnarray}
\nonumber S_L^x-S_R^x &=& {1\over{2 \sqrt{2}}} \Bigl{(} \ket{S_{11}} \bra{T_{11}^-} - \ket{S_{11}}\bra{T_{11}^+} \ +
h.c.\Bigr{)}\\
\nonumber S_L^y-S_R^y &=& {1\over{2 \sqrt{2}}} \Bigl{(} i\ket{S_{11}}\bra{T_{11}^-} - i\ket{S_{11}}\bra{T_{11}^+} \ +
h.c.\Bigr{)}\\
S_L^z-S_R^z &=& {1\over{2}} \Bigl{(} \ket{S_{11}}\bra{T_{11}^0} + \ket{T_{11}^{0}}\bra{S_{11}}\Bigr{)} \label{E_STcoupling}.
\end{eqnarray}
The first two expressions reveal that the inhomogeneous field in the transverse plane $\Delta B_N^x$, $\Delta B_N^y$ mixes the $\ket{T_{11}^+}$ and
$\ket{T_{11}^-}$ states with the $\ket{S_{11}}$. The longitudinal component $\Delta B_N^z$ mixes $\ket{T_{11}^0}$ with $\ket{S_{11}}$ (third
expression). The degree of mixing between two states will depend strongly on the energy difference between them \cite{koppensscience}. In the case of
$g \mu_B B_{ext},t<g \mu_B \sqrt{\langle\Delta{B_N^2}\rangle}$, the three triplet states are close in energy to the $\ket{S_{11}}$ state. Their
intermixing will be strong, lifting spin blockade. For $g \mu_B B_{ext}\gg t,g \mu_B \sqrt{\langle\Delta B_N^2\rangle}$ the $\ket{T^{+}_{11}}$ and
$\ket{T^{-}_{11}}$ states are split off in energy by an amount of $g\mu_{B}B_{ext}$. Consequently the perturbation of these states caused by the
nuclei will be small. Although the $\ket{T_{11}^{0}}$ remains mixed with the $\ket{S_{11}}$ state, the occupation of one of the two split-off triplet
states can block the flow through the system.

The effect of nuclear mixing is shown in Fig. \ref{F_STmixing} \cite{koppensscience}. The observed current flow through the system is typically in
the order of a few hundreds of fA (Fig. \ref{F_STmixing}a). At zero field, where the mixing is strongest, the current flow is largest. Increasing the
field gradually restores spin blockade. Fig. \ref{F_STmixing}b shows the energy levels for zero and finite external field. The theoretical
calculations of the nuclear-spin mediated current flow (obtained from a master equation approach) are discussed in references
\cite{jouravlevprl,platerocondmat}.

\subsection{Oscillating magnetic field and rate equations}

So far, we have seen that the occurrence of transitions between singlet and triplet spin states are detectable as a small current in the spin
blockade regime. We will now discuss how this lifting of spin blockade can also be used to detect single spin rotations, induced via electron spin
resonance. The basic idea is the following. The basic idea is the following. If the system is blocked in e.g. $\ketup\ketup$, and the driving field
rotates e.g. the left spin, then transitions are induced to the state $\ketdown \ketup$. This state contains a singlet component and therefore a
probability for the electron to move to the right dot and right lead. Inducing single spin rotations can therefore lift spin blockade.

However, together with the driving field, the spin transitions are much more complicated due to the interplay of different processes: spin resonance
of the two spins, interaction with the nuclear fields, spin state hybridization by coherent dot coupling and inelastic transitions from the S(1,1)
state to the S(0,2) state. In order to understand the interplay of these processes, we will first model the system with a time-dependent Hamiltonian
and a density matrix approach. Next, we will discuss the physical interpretation of the simulation results.

The Hamiltonian now also contains a term with an oscillating magnetic field in the x-direction with amplitude $B_{ac}$
\begin{equation}
H_{ac}(t) = g\mu_{B}B_{ac}\sin(\omega \tau)\cdot (S_{L}^x+S_{R}^x).
\end{equation}
We assume that $B_{ac}$ is equal in both dots, which is a reasonable approximation in the experiment (from simulations we find that the difference of
$B_{ac}$ is 20\% at most \cite{koppensnature}). We assume $B_{ext}\gg B_N,B_{ac}$, which allows application of the rotating wave approximation
\cite{Poole}. Therefore, we will define $B_1\equiv\frac{1} 2 B_{ac}$, which is in the rotating frame the relevant driving field for the ESR process.

In order to study the effect of ESR and the nuclear fields that are involved in the transport cycle, we will construct rate equations that include
the unitary evolution of the spins in the dots governed by the time-dependent Hamiltonian. This approach is based on the model of reference
\cite{jouravlevprl}, where the Hamiltonian contained only time-independent terms. Seven states are involved in the transport cycle, namely the three
(1,1) triplets $\ket{T^{i}_{11}}$, the double and single dot singlet states $\ket{S_{11}}$ and $\ket{S_{02}}$ and the two (0,1) states
$\ket{\uparrow_{01}}$ and $\ket{\downarrow_{01}}$, making the density operator a $7\times7$ matrix. The rate equations based on the time-independent
Hamiltonian are given in \cite{jouravlevprl}. These are constructed from the term that gives the unitary evolution of the system governed by the
Hamiltonian ($H=H_0+H_{ac}$) $d\hat{\rho}_{k}/d\tau = -\frac{i}{\hbar} \bra{k} [H,\hat{\rho}] \ket{k}$, together with terms that account for
incoherent tunnelling processes between the states. The rate equations for the diagonal elements are given by
\begin{eqnarray}\label{E_ratediag}
\nonumber {d\hat{\rho}_{T^+_{11}}\over{d\tau} }&=&  -\frac{i}{\hbar} \bra{T^+_{11}}[H,\hat{\rho}]\ket{T^{+}_{11}} + {\Gamma_{L}\over{2}}
\hat{\rho}_{\uparrow_{01}}\\
\nonumber {d\hat{\rho}_{T^-_{11}}\over{d\tau} }&=&  -\frac{i}{\hbar} \bra{T^-_{11}}[H,\hat{\rho}]\ket{T^{-}_{11}} + {\Gamma_{L}\over{2}}
\hat{\rho}_{\downarrow_{01}}\\
\nonumber {d\hat{\rho}_{T^0_{11}}\over{d\tau} }&=&  -\frac{i}{\hbar} \bra{T^0_{11}}[H,\hat{\rho}]\ket{T^{0}_{11}} + {\Gamma_{L}\over{4}}
\bigl{(}\hat{\rho}_{\uparrow_{01}} + \hat{\rho}_{\downarrow_{01}}\bigr{)}\\
\nonumber {d\hat{\rho}_{S_{11}}\over{d\tau} }&=&  -\frac{i}{\hbar} \bra{S_{11}}[H,\hat{\rho}]\ket{S_{11}} + {\Gamma_{L}\over{4}}
\bigl{(}\hat{\rho}_{\uparrow_{01}} + \hat{\rho}_{\downarrow_{01}}\bigr{)} -
\Gamma_{in} \hat{\rho}_{S_{11}}\\
\nonumber {d\hat{\rho}_{S_{02}}\over{d\tau} }&=&  -\frac{i}{\hbar} \bra{S_{02}}[H,\hat{\rho}]\ket{S_{02}} + \Gamma_{in}
\hat{\rho}_{S_{11}} - \Gamma_{R}\hat{\rho}_{S_{02}}\\
\nonumber {d\hat{\rho}_{\uparrow_{01}}\over{d\tau} }&=& + {\Gamma_{R}\over{2}}\hat{\rho}_{S_{02}} -
\Gamma_{L}\hat{\rho}_{\uparrow_{01}}\\
{d\hat{\rho}_{\downarrow_{01}}\over{d\tau} }&=& + {\Gamma_{R}\over{2}}\hat{\rho}_{S_{02}} - \Gamma_{L}\hat{\rho}_{\downarrow_{01}}
\end{eqnarray}

\begin{figure}[b]
     \includegraphics[scale=1.3]{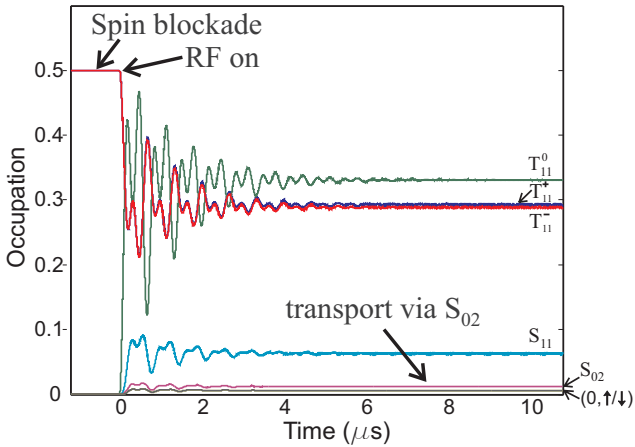}
     \centering
     \caption{ Time evolution of the diagonal elements of the density matrix for one particular nuclear configuration. Parameters: $\hbar\omega=g\mu_B 100$mT, $B_{ext}=$100 mT, $B_{N,x,y,z}^L=$(0,0,2.2) mT, $B_{N,x,y,z}^R=$(0,0,0), $B_1=1.3$ mT, $\Gamma_L=73$ MHz, $\Gamma_R=73$ MHz, $\hbar\Gamma_{in}=g\mu_B B_{N,z}^L$ and $\Delta_{LR}$=200$\mu$eV, $t$=0.3 $\mu$eV.}
     \label{densityevolution}
\end{figure}

The rate equations for the off-diagonal elements are given by
\begin{equation} \label{E_rateoffdiag}
{d\hat{\rho}_{jk}\over{d\tau} } = -\frac{i}{\hbar} \bra{j} [H,\hat{\rho}] \ket{k} - {1\over{2}}\bigl{(}\Gamma^{j} + \Gamma^{k} \bigr{)}
\hat{\rho}_{jk}
\end{equation}
where the indices $j,k \in \bigl{\{} T_{11}^{i},S_{11},S_{02}, \uparrow_{01}, \downarrow_{01} \bigr{\}}$ label the states available to the system.
The tunneling/projection rates $\Gamma^j$ equal $\Gamma_{in}$ and $\Gamma_{R}$ for the $\ket{S_{11}}$ and $\ket{S_{02}}$ states respectively, and
equal zero for the other 5 states. The first term on the right-hand side describes the unitary evolution of the system, while the second term
describes a loss of coherence due to the finite lifetime of the singlet states. This is the first source of decoherence in our model. The second one
is the inhomogeneous broadening due to the interaction with the nuclear system.  We do not consider other sources of decoherence, as they are
expected to occur on much larger timescales.

Because we added a time-dependent term to the Hamiltonian (the oscillating field), we numerically calculate the time evolution of $\hat{\rho}(t)$,
treating the Hamiltonian as stationary on the timescale $\Delta \tau \ll 2\pi /\omega$. To reduce the simulation time, we use the steady state
solution $\hat{\rho}_{\tau \rightarrow \infty}$ in the absence of the oscillating magnetic field as the initial state $\hat{\rho}(\tau=0)$ for the
time evolution. At $\tau=0$ the oscillating field is turned on and the system evolves towards a dynamic equilibrium on a timescale set by the inverse
of the slowest tunnelling rate $\Gamma$. This new equilibrium distribution of populations is used to calculate the current flow, which is
proportional to the occupation of the $\ket{S_{02}}$ state ($I = e \Gamma_{R} \hat{\rho}_{S_{02}}$). An example of the time evolution of the density
matrix elements is shown in Fig. \ref{densityevolution}. The figure clearly reveals that the blockade is lifted when the oscillating field is
applied. This is visible as an increase of the occupation of the $\ket{S_{02}}$ state.

In order to simulate the measured current flow we have to consider the fact that the measurements are taken with a sampling rate of ~1 Hz. As the
timescale of the nuclear dynamics is believed to be much faster than 1 Hz \cite{coishnonmarkovian,Sousa,Paget77}, we expect each datapoint to be an
integration of the response over many configurations of the nuclei. The effect of the evolving nuclear system is included in the calculations by
averaging the different values of the (calculated) current flow obtained for each frozen configuration. These configurations are randomly sampled
from a gaussian distribution of nuclear fields in the left and right dot (similar as in \cite{jouravlevprl}). Because the electron in the two dots
interact with different nuclear spins, the isotropic gaussian distributions in the two dots are uncorrelated, such that $\sqrt{\langle \Delta B_N^2
\rangle}=\sqrt{2} \sqrt{ \langle B_{N}^2 \rangle}$ and $\langle B_{N,x}^2 \rangle =
 \langle B_{N,y}^2 \rangle = \langle B_{N,z}^2 \rangle$. For the sake of convenience we define
 \begin{equation}
\sigma_N=\sqrt{\langle B_{N}^2 \rangle} \quad \text{and} \quad \sigma_{N,z}=\sqrt{\langle B_{N,z}^2 \rangle}=\sqrt{\frac{1}3\langle B_{N}^2 \rangle}.
\end{equation}

\subsection{Simulation results and physical picture}

\begin{figure}[]
     \includegraphics[scale=1]{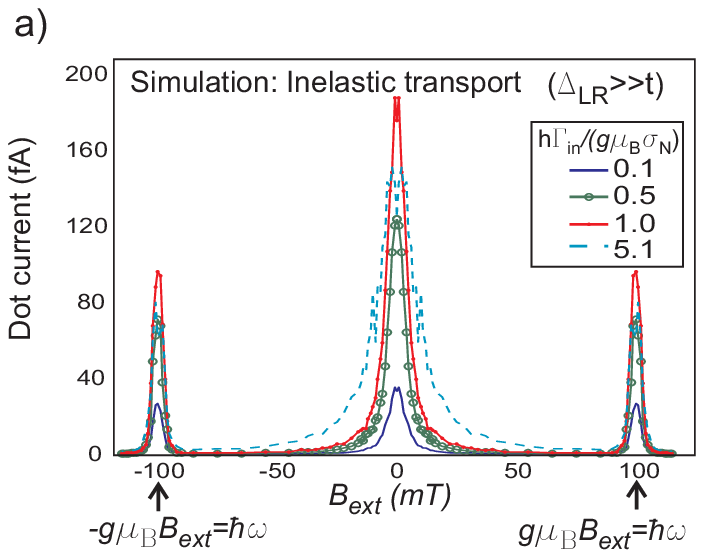}
          \includegraphics[scale=1]{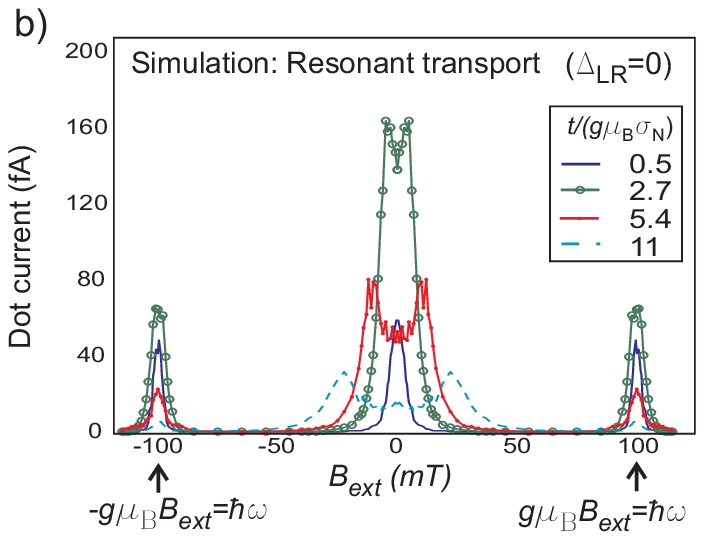}
\includegraphics[scale=1]{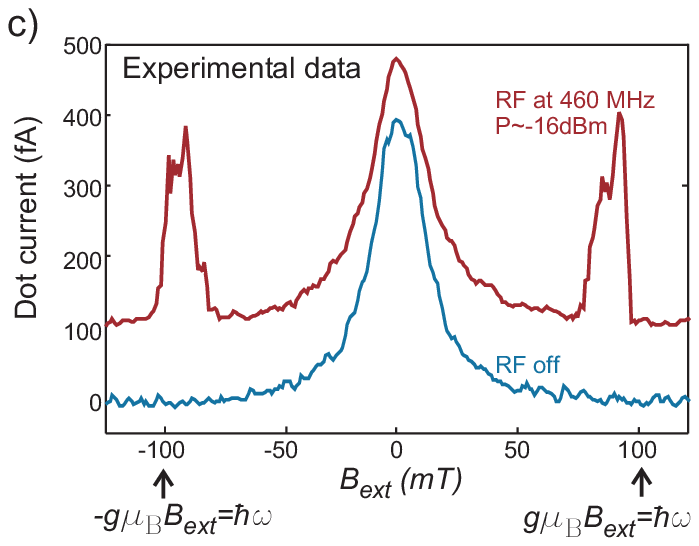}
     \centering
     \caption{ a). Calculated average current flow in the inelastic transport regime. Parameters: $\hbar\omega=g\mu_B 100$mT, $B_{ext}=$100 mT, $\sigma_{N}=$2.2mT, $B_1=1.3$mT, $\Gamma_{L,R}=73$ MHz, $t$=0.3 $\mu$eV and $\Delta_{LR}=200$ $\mu$eV. Results are similar for any value for $t$, provided that $\Delta_{LR} \gg t$. b) Calculated average current flow in the resonant transport regime at zero detuning for different values of $t$. Parameters: $\hbar\omega=g \mu_B 100$mT, $\sigma_{N}=$2.2mT, $B_1=1.3$mT, $\Gamma_{L,R}=73$ MHz, $\Gamma_{in}=0$ and $\Delta_{LR}=0$. Averaged over 400 nuclear configurations for $t/(g\mu_B\sigma_N)>0.5$ and 60 configurations for $t/(g\mu_B\sigma_N)=0.5$. Simulation carried out for positive magnetic fields only; values shown for negative fields are equal to results obtained for positive field. c) Experimental data from Ref \cite{koppensnature} with (curve offset by 100 fA for clarity) and without oscillating magnetic field. The frequency of the oscillating magnetic field is 460 MHz and the applied power is $~$-16dBm. Simulation carried out for positive magnetic fields only; values shown for negative fields are equal to results obtained for positive field.}
     \label{leakagecalc}
\end{figure}

An example of the calculated (average) current flow as a function of $B_{ext}$ (Fig. \ref{leakagecalc}a,b) shows a (split) peak around zero magnetic
field and two satellite peaks for $B_{ext}=\pm\hbar \omega /(g \mu_B)$, where the spin resonance condition is satisfied. This (split) peak at
$B_{ext}=0$ is due to singlet-triplet mixing by the inhomogeneous nuclear field, and the splitting depends on the tunnel coupling, similar as the
observations in  \cite{koppensscience}. The response from the induced spin flips via the driving field is visible for the both inelastic and resonant
transport regime, and the current flow has comparable magnitude to the peak at $B_{ext}=0$. The satellite peaks are also visible in the experimental
data from \cite{koppensnature} (also shown here in Fig. $\ref{leakagecalc}$), although the shape and width of the satellite peaks are different, as
we will discuss later.

We want to stress that the ESR satellite peaks only appear when an inhomogeneous nuclear field is present in the simulations. In other words, for
$\Delta B_N=0$ and $B_1$ equal in both dots, spin rotations are induced in both dots at the same time and at the same rate. Starting, for example,
from the state $\ket{T_{11}^+}=\ket{\! \uparrow \uparrow}$ transitions are induced to the state $\ket{\! \downarrow \downarrow}$ via the intermediate
state $\ket{\!\uparrow+\downarrow} \ket{\!\uparrow+\downarrow}/\sqrt{2}=(\ket{T_{11}^+}+\ket{T_{11}^-}+2\ket{T_{11}^0})/\sqrt{2}$. No mixing with the
singlet state takes place (the evolution is in the triplet-subspace) and no current will therefore flow.

The ESR sattelite peaks are visible for both resonant and inelastic transport regime (Figs. \ref{leakagecalc}a,b). For the resonant transport regime,
we see that for $t/\sigma_N<5$ the sattelite peak increases in height when increasing $t$, simply because the coupling between the two singlet states
increases. However, further increasing $t$ reduces the signal, and this is because the exchange splitting then plays a more important role. Namely,
increasing the exchange splitting reduces the mixing between the $T_{11}^0$ state with the hybridized singlet state by the nuclear field gradient.
This mixing is a crucial element for detecting the induced rotations of one of the two electron spins. In the inelastic transport regime, this
exchange splitting is negligibly small and therefore the height of the sattelite peak depends only on $\Gamma_{in}$ and the driving field $B_1$.

\begin{figure}[]
     \includegraphics[scale=0.9]{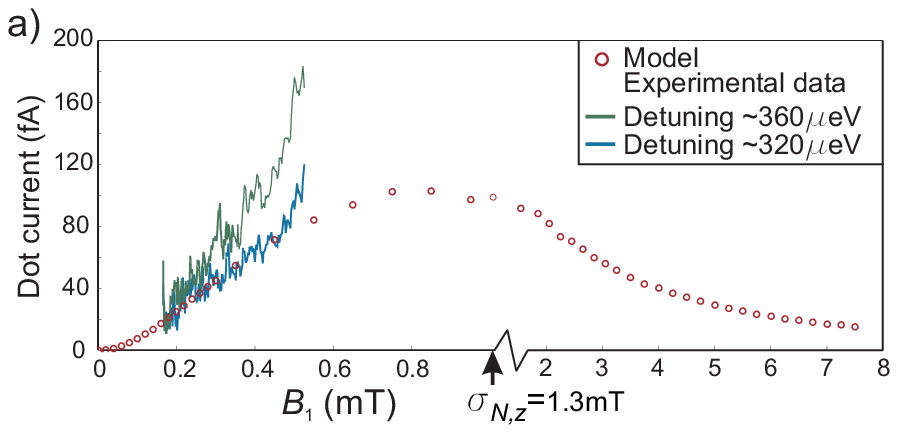}
          \includegraphics[scale=1]{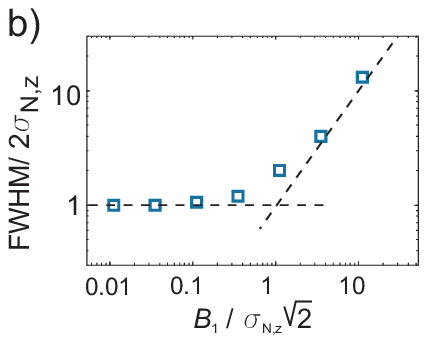}
   %  \centering
     \caption{Height and width of the ESR satellite peak. a) Circles: calculated ESR peak height as a function of driving amplitude $B_{1}$. Parameters: $\hbar\omega=g\mu_B 100$mT, $B_{ext}=$100 mT, $\sigma_{N}=$2.2mT, $\Gamma_{L,R}=73$ MHz, $t$=0.3 $\mu$eV, $\hbar\Gamma_{in}=g\mu_B \sigma_{N}$  and $\Delta_{LR}=200$ $\mu$eV. Lines are the current measurements for 2 different values of $\Delta_{LR}$. The measurements show time-dependent (telegraph type) behavior. Therefore, the curves are obtained by repeating sweeps of $B_1$ and then selecting the largest current value for each value of $B_1$. b) Calculated width of the ESR satellite peaks as a function of $B_1$. For small ESR power the peak is broadened by the random nuclear fluctuations, at high powers it is broadened by $B_{1}$.}
     \label{F_ESRSTsat}
\end{figure}

A study of the height of the satellite peak as a function of $B_1$ reveals a non-monotonous behaviour, which can be seen in Fig. \ref{F_ESRSTsat}a.
The physical picture behind this behavior is most easily sketched by distinguishing three regimes:
\begin{enumerate}
    \item
     For $B_1<\sigma_{N,z}$, for most of the nuclear configurations the spin in at most one of the two dots is on resonance, so spins are flipped in either the left or right dot.  In that case transitions are induced from e.g. $\ket{\!\uparrow \uparrow}$ to $\ket{\!\uparrow \downarrow}=\ket{\!S_{11}}+\ket{\!T_{11}^0}$ or
$\ket{\!\downarrow\uparrow}=\ket{\!S_{11}}-\ket{\!T_{11}^0}$. The resulting current flow initially increases quadratically with $B_1$, as one would
normally expect (Fig. \ref{F_ESRSTsat}a).

    \item  For $B_1\gg\sigma_{N,z}$, for most of the nuclear configurations two spins are rotated simultaneously due to power broadening of the Rabi resonance. The stronger $B_1$, the more the transitions occur only in the triplet subspace (the driving field $B_1$ that rotates two spins dominates the $S-T_0$ mixing by the nuclear spins). As a result, the current decreases for increasing $B_1$.

\item
If $B_1\sim\sigma_{N,z}$ the situation is more complex because both processes (rotation of 2 spins simultaneously \textit{and} transitions from
$T_{11}^0$ to $S_{11}$) are effective. We find that if both processes occur with comparable rates, the overall transition rate to the singlet state
is highest. This is the reason why the current has a maximum at $B_1\approx\sigma_{N,z}$ (Fig. \ref{F_ESRSTsat}a).
\end{enumerate}

The experimental data of the ESR satellite peak height (normalized by the zero-field current flow) for two different values of $\Delta_{LR}$ are
shown in Fig. \ref{F_ESRSTsat}a. In order to compare the experimental results with the model we have estimated the rate $\Gamma_{in}$ from the
measured current flow at $B_{ext}=0$ (we found similar values for both curves). The agreement of the experimental data with the model is reasonable,
as it shows the expected quadratic increase with $B_1$, as well as a comparable peak height. However, we see that variations of the level detuning
$\Delta_{LR}$ can result in considerable differences of the measured ESR peak height. We have two possible explanations for the deviations of the
experimental data with the model. First, we have found experimental signatures of dynamic nuclear polarization when the ESR resonance condition was
fulfilled. We expect that this is due to feedback of the electron transport on the nuclear spins (similar to that discussed in
\cite{onoprl,platerocondmat,rudnercondmat}), although the exact processes are not (yet) fully understood. Second, unwanted electric fields affect the
electron tunnelling processes, but are not taken into account in the model. We expect that these electric fields will not change the location and
width of the ESR sattelite peaks because this field does not couple the spin states. It is however possible that the height of the satellite peak is
altered by the electric field because if can affect the coupling between the S(0,2) with the S(1,1) state.

Finally, we discuss the width of the ESR satellite peak (Fig. \ref{F_ESRSTsat}b). If the inelastic tunnelling process between the dots (with rate
$\Gamma_{in}$) and $B_1$ are both smaller than $\sigma_{N,z}$, the ESR peak (obtained from simulations) is broadened by the statistical fluctuations
of the effective nuclear field. For high $B_1$, the width approaches asymptotically the line with slope 1 (see Fig. \ref{F_ESRSTsat}b). In this
regime, the peak is broadened by the RF amplitude $B_1$. In the experiment \cite{koppensnature}, the shape of the satellite peak was different (flat
on top with sharp edges) than expected from the model. Furthermore, the FWHM was larger than expected from just $\sigma_{N,z}$. We attribute this to
feedback of the ESR-induced current flow on the nuclear spin bath. As a result, a clear FWHM increase with $B_1$ could not be observed.

It should be noted that in the simulation the central peak is broader than the satellite peaks. From studying the influence of various parameters in
the model, we conclude that the greater width of the central peak is caused by the tranverse nuclear field fluctuations ($B_{N,x}$ and $B_{N,y}$),
which broaden the central peak but not the ESR satellite peaks.

We conclude that the model discussed here qualitatively agrees with the main features that were observed in the double dot transport measurements
that aims at detecting (continuous wave) ESR of a single electron spin. The details of the ESR satellite peak height and width do not agree
quantitatively with the model. We believe these deviations can be attributed to unwanted electric fields and feedback of the electron transport on
the nuclear spin polarization. Improving the understanding of these feedback mechanisms remains interesting for future investigation as it might
point towards a direction to mitigate the decoherence of the electron spin \cite{klauserprb,jouravlevprl}.

\begin{acknowledgments}
This study was supported by the Dutch Organization for Fundamental Research on Matter (FOM), the Netherlands Organization for Scientific Research
(NWO) and the Defense Advanced Research Projects Agency Quantum Information Science and Technology program.
\end{acknowledgments}

%\newpage

%\bibliographystyle{apssamp}
%\bibliography{Bib_ICPS}
%\bibliography{Bib_Koppens}

\newpage
\newcommand{\noopsort}[1]{} \newcommand{\printfirst}[2]{#1}
  \newcommand{\singleletter}[1]{#1} \newcommand{\switchargs}[2]{#2#1}

\newpage
  \baselineskip24pt

\end{document}